\documentclass[showpacs,aps,floatfix,prd,preprintnumbers]{revtex4}
\usepackage{amssymb}
\usepackage{amsmath}
\usepackage{epstopdf}
\usepackage{hyperref}
\usepackage{capt-of}
\usepackage{graphicx}
\usepackage{dcolumn}
\usepackage{bm}
\usepackage{booktabs}
\usepackage{multirow}
\usepackage{mathtools}
\usepackage{xcolor}
\usepackage{orcidlink}
\RequirePackage{float}

\begin{document}

\title{$f(Q,T)$ gravity models with observational constraints}

\author{Simran Arora\orcidlink{0000-0003-0326-8945}}
\email{dawrasimran27@gmail.com}
\affiliation{Department of Mathematics, Birla Institute of Technology and
Science-Pilani,\\ Hyderabad Campus, Hyderabad-500078, India.}
\author{S. K. J. Pacif\orcidlink{0000-0003-0951-414X}}
\email{shibesh.math@gmail.com}
\affiliation{Department of Mathematics, School of Advanced Sciences, Vellore
Institute of Technology, Vellore 632014, Tamil Nadu, India}
\author{Snehasish Bhattacharjee\orcidlink{0000-0002-7350-7043}}
\email{snehasish.bhattacharjee.666@gmail.com}
\affiliation{Department of Astronomy, Osmania University, Hyderabad-500007, India.}
\author{P.K. Sahoo\orcidlink{0000-0003-2130-8832}}
\email{pksahoo@hyderabad.bits-pilani.ac.in}
\affiliation{Department of Mathematics,\\ Birla Institute of Technology and
Science-Pilani, Hyderabad Campus, Hyderabad-500078, India.}
\date{\today}

\begin{abstract}
The paper presents late time cosmology in $f(Q,T)$ gravity where the dark
energy is purely geometric in nature. We start by employing a well motivated 
$f(Q,T)$ gravity model, $f(Q,T)=mQ^{n}+bT$ where $m,n$ and $b$ are model
parameters. Additionally we also assume the universe to be dominated by
pressure-less matter which yields a power law type scale factor of the form $%
a(t)=c_{2}(At+c_{1})^{\frac{1}{A}}$, where $A=\dfrac{3(8\pi +b)}{n(16\pi +3b)%
}$ and $c_{1}$ \& $c_{2}$ are just integration constants. To investigate the
cosmological viability of the model, constraints on the model parameters
were imposed from the updated 57 points of Hubble data sets and 580 points
of union 2.1 compilation supernovae data sets. We have thoroughly
investigated the nature of geometrical dark energy mimicked by the
parametrization of $f(Q,T)=mQ^{n}+bT$ with the assistance of statefinder
diagnostic in $\{s,r\}$ and $\{q,r\}$ planes and also performed the $Om$
-diagnostic analysis. The present analysis makes it clear-cut that $f(Q,T)$
gravity can be promising in addressing the current cosmic acceleration and
therefore a suitable alternative to the dark energy problem. Further studies
in other cosmological areas are therefore encouraging to further investigate
the viability of $f(Q,T)$ gravity.
\end{abstract}

\maketitle


\section{Introduction}

\label{I}

Cosmological observations indicate that our Universe is going through a
phase of accelerated expansion \cite{Riess,Perlmutter}. With regard to the
theoretical predictions and some leading surveys in observational
perspective indicates the presence of a curious form of energy with high
negative pressure and increasing density. The cosmological entity
responsible for the acceleration should account for almost three-quarters of
the such curious energy budget of the universe. In addition to that, this
entity should also be able to create an anti-gravity effect permeating
throughout the entire observable universe. Since normal baryonic matter do
not posses such an equation of state, let alone to account for such
alion-share of the energy budget of the universe, several alternate
scenarios have been proposed and investigated \cite{alternate}.

Interestingly, such an accelerated expansion is possible with normal
baryonic matter if the gravitational forces felt by cosmic objects turns out
to be different than that proposed by general relativity \cite{cosmoft}.
Such scenarios are built upon modified theories of gravity in which the
geometrical sector of the field equations undergoes a subtle modifications
without altering the matter sector. Extended theories of gravity such as $%
f(R)$ gravity, $f(G)$ gravity, $f(R,T)$ gravity, etc. are widely employed in
modern cosmology (For a recent review on modified gravity see \cite{review}.
Also see \cite{snehasish} for some interesting cosmological applications of
modified gravity) to address the late-time acceleration and other
shortcomings of $\Lambda $CDM cosmologies.

$f(Q,T)$ gravity is a recently proposed extended theory of gravity in which
the Lagrangian density of the gravitational field is a function of the
non-metricity $Q$ and trace of energy momentum tensor $T$ \cite{Yixin}. In 
\cite{harko/ft}, the authors studied inflationary and late time cosmology in 
$f(\mathcal{T},T)$ gravity where $\mathcal{T}$ and $T$ are torsion and trace
of energy momentum tensor. A very recent study of Yixin \cite{Yixin Xu}
shows the non-minimal coupling between $Q$ and $T$ taking into account some
different class of cosmological models with some specific functional forms
of $f(Q,T)$. Different forms of $f(Q,T)$ can result in obtaining a large
variety of cosmological evolution including the deceleration and
acceleration expansions. Also the study by Bhattacharjee \cite{epjc}, $%
f(Q,T) $ gravity was found to yield excellent theoretical estimates of
baryon-to-entropy ratios and therefore could solve the puzzle of
over-abundance of matter over anti-matter. With that in mind, we seek out to
investigate the cosmological viability of $f(Q,T)$ gravity in sufficing the
conundrum of late-time acceleration without incorporating dark energy.

The manuscript is organized as follows: The present cosmological scenario is
studied and introduced in section \ref{I}. In Section \ref{II}, we present
an overview of $f(Q,T)$ gravity. In Section \ref{III}, we present the
cosmological model employed in the work with some model parameters and
derived some physical parameters. In accordance, the non-parametric method
is sometimes more beneficial as the evolution of our universe can be found
directly from the observational data. Therefore, in section \ref{IV}, we
constrain the model parameters using Hubble ($H(z)$) data-sets and
Supernovae ($SN$) data-sets. In Section \ref{V}, we present some geometrical
diagnostics which results in distinction between various dark energy models
and $\Lambda $CDM. We present the behavior of energy density in Section \ref%
{VI}. In Section \ref{VII}, a special case is discussed. Finally, in Section %
\ref{VIII}, we present our results and conclusions.

\section{Overview of $f(Q,T)$ Gravity}

\label{II}

The action in $f(Q,T)$ Gravity is given as \cite{Yixin},

\begin{equation}  \label{1}
S=\int \left( \frac{1}{16\pi} f(Q,T) + L_{m}\right)d^4x \sqrt{-g} .
\end{equation}
where $f$ is an arbitrary function of the non-metricity $Q$ and the trace of
the matter-energy-momentum tensor $T$, $L_{m}$ represents the matter
Lagrangian and $g = det(g_{\mu\nu}$ and

\begin{equation}  \label{2}
Q\equiv
-g^{\mu\nu}(L_{\beta\mu}^{\alpha}L_{\nu\alpha}^{\beta}-L_{\beta\alpha}^{%
\alpha}L_{\mu\nu}^{\beta}).
\end{equation}
where $L^{\alpha}_{\beta\gamma}$ is the deformation tensor given by, 
\begin{equation}  \label{3}
L^{\alpha}_{\beta\gamma}=-\frac{1}{2}g^{\alpha\lambda}(\nabla_{\gamma}g_{%
\beta\lambda}+\nabla_{\beta}g_{\lambda\gamma}-\nabla_{\lambda}g_{\beta
\gamma}).
\end{equation}

The non-metricity $Q$ and trace of energy momentum tensor $T$ are defined
respectively as 
\begin{equation}  \label{4}
Q_{\alpha} \equiv {Q_{\alpha}^{\,\, \mu}}_{\, \mu}, \hspace{0.15in}
T_{\mu\nu}= -\frac{2}{\sqrt{-g}}\frac{\delta(\sqrt{-g}L_{m}}{\delta
g^{\mu\nu}}
\end{equation}
and 
\begin{equation}  \label{5}
\Theta_{\mu \nu}= g^{\alpha \beta} \frac{\delta T_{\alpha \beta}}{\delta
g^{\mu \nu}}.
\end{equation}

The variation of the gravitational action (\ref{1}) leads to the following
field equation

\begin{equation}  \label{6}
8\pi T_{\mu \nu}= -\frac{2}{\sqrt{-g}}\nabla_{\alpha}(f_{Q}\sqrt{-g}%
P^{\alpha}_{\, \mu\nu}-\frac{1}{2}fg_{\mu \nu}+ f_{T}(T_{\mu\nu}+\Theta_{\mu
\nu})-f_{Q}(P_{\mu\alpha\beta}Q_{\nu}^{\,\,\alpha \beta}-2Q^{\alpha
\beta}_{\, \, \, \mu}P_{\alpha\beta\nu}).
\end{equation}
where $P^{\alpha}_{\,\,\, \mu \nu}$ is the super-potential of the model as
mentioned in \cite{Yixin}.

We now assume a flat FLRW metric as, 
\begin{equation}  \label{7}
ds^{2}=-N^{2}(t)dt^{2}+a(t)^{2}(dx^{2}+dy^{2}+dz^{2}),
\end{equation}
where $a(t)$ is the scale factor and $N(t)$ the Lapse function. We consider
the case when $N(t)=1$ i.e. the case of the standard FLRW geometry. Thus we
have $Q=6H^{2}$ and the generalized Friedmann equations are, 
\begin{equation}  \label{8}
8\pi \rho =\frac{f}{2}-6FH^{2}-\frac{2\overline{G}}{1+\overline{G}}(\dot{F}%
H+F\dot{H}).
\end{equation}
and 
\begin{equation}  \label{9}
8\pi p=-\frac{f}{2}+6FH^{2}+2(\dot{F}H+F\dot{H}).
\end{equation}

where dot represents derivative with respect to time and $F= f_{Q}$ and $8
\pi \overline{G}=f_{T}$ represent differentiation with respect to $Q$ and $T$
respectively. Also note that $F= f_{Q}= m n Q^{n-1}$ and $8 \pi\overline{G}%
=f_{T}=b$.

Combining equations \eqref{8} \& \eqref{9} we can get 
\begin{equation}  \label{10}
\dot{H}+\dfrac{\dot{F}}{F} H= \dfrac{4\pi}{F}(1+\overline{G})(\rho+p).
\end{equation}

For generality we assume that the cosmological matter satisfies an equation
of state of the form $p=(\gamma -1)\rho $, where $\gamma $ is a constant,
and $1\leq \gamma \leq 2$. Solving Eqs. (\ref{8}) and (\ref{10}), we
obtained the energy density $\rho $ as

\begin{equation}  \label{11}
\rho= \dfrac{f-12F H^{2}}{16\pi(1+\gamma \overline{G})}.
\end{equation}

\section{Cosmological model with $f(Q,T)= m Q^{n}+b T$}\label{III}

In this section, we shall consider the functional form of $f(Q,T)$ as \cite%
{Yixin} 
\begin{equation}  \label{12}
f(Q,T)=mQ^{n}+bT,
\end{equation}
where $m$, $b$ and $n$ are model parameters.

We find the solution for zero pressure (dust matter) for which, we have $%
\gamma =1$ and Eq. (\ref{11}) together with the $f(Q,T)$ function given in (%
\ref{12}), will reduce to, 
\begin{equation}  \label{13}
\rho =\dfrac{m6^{n}H^{2n}(1-2n)}{16\pi +3b}\text{.}
\end{equation}

Now, the dynamical equation Eq. (\ref{10}) describing the dynamics of the
model reads as, 
\begin{equation}
\dot{H}+\dfrac{3(8\pi +b)}{n(16\pi +3b)}H^{2}=0\text{,}  \label{14}
\end{equation}
which readily integrated to give the time evolution of the Hubble parameter $%
H(t)$ as, 
\begin{equation}  \label{15}
H(t)=\frac{1}{At+c_{1}}\text{, where }A=\dfrac{3(8\pi +b)}{n(16\pi +3b)}
\end{equation}
and $c_{1}$ is a constant of integration. From, equation (\ref{15}), we
obtain the explicit form of scale factor as a simple power law type solution
given by, 
\begin{equation}  \label{16}
a(t)=c_{2}(At+c_{1})^{\frac{1}{A}}\text{.}
\end{equation}
where $c_{2}$ is another constant of integration. As, we are dealing with
zero pressure matter and trying to explain the present cosmic acceleration,
we shall express all the above cosmological parameters in terms of redshift $%
z$, defined by $z=\frac{a_{0}}{a}-1$, where $a_{0}$ is the present value (at
time $t=t_{0}$) of the scale factor. Also, we will consider the normalized
value $a_{0}=c_{2}(At_{0}+c_{1})^{\frac{1}{A}}=1$ for which the $t$-$z$
relationship will be established as,

\begin{equation}  \label{17}
t(z)=-\frac{c_{1}}{A}+\frac{1}{A}\left[ c_{2}(1+z)\right] ^{-A}\text{.}
\end{equation}

Using the $t$-$z$ relation in Eq. (\ref{17}), the Hubble parameter in terms
of $z$ can be written as, 
\begin{equation}  \label{Hz}
H(z)=H_{0}(1+z)^{A}=H_{0}(1+z)^{\frac{3(8\pi +b)}{n(16\pi +3b)}}\text{,}
\end{equation}
containing only two model parameters $n$ and $b$. The deceleration parameter 
$q=-1-\frac{\dot{H}}{H^{2}}$comes out to be, 
\begin{equation}
q(t)=-1+\dfrac{3(8\pi +b)}{n(16\pi +3b)}\text{,}  \label{q}
\end{equation}
which is a constant as expected due to power-law type expansion of the model.

\section{Parameters of the model \& observational constraints}\label{IV}

\begin{figure}[H]
\begin{center}
$\includegraphics[width=3.0 in, height=2.5
		in]{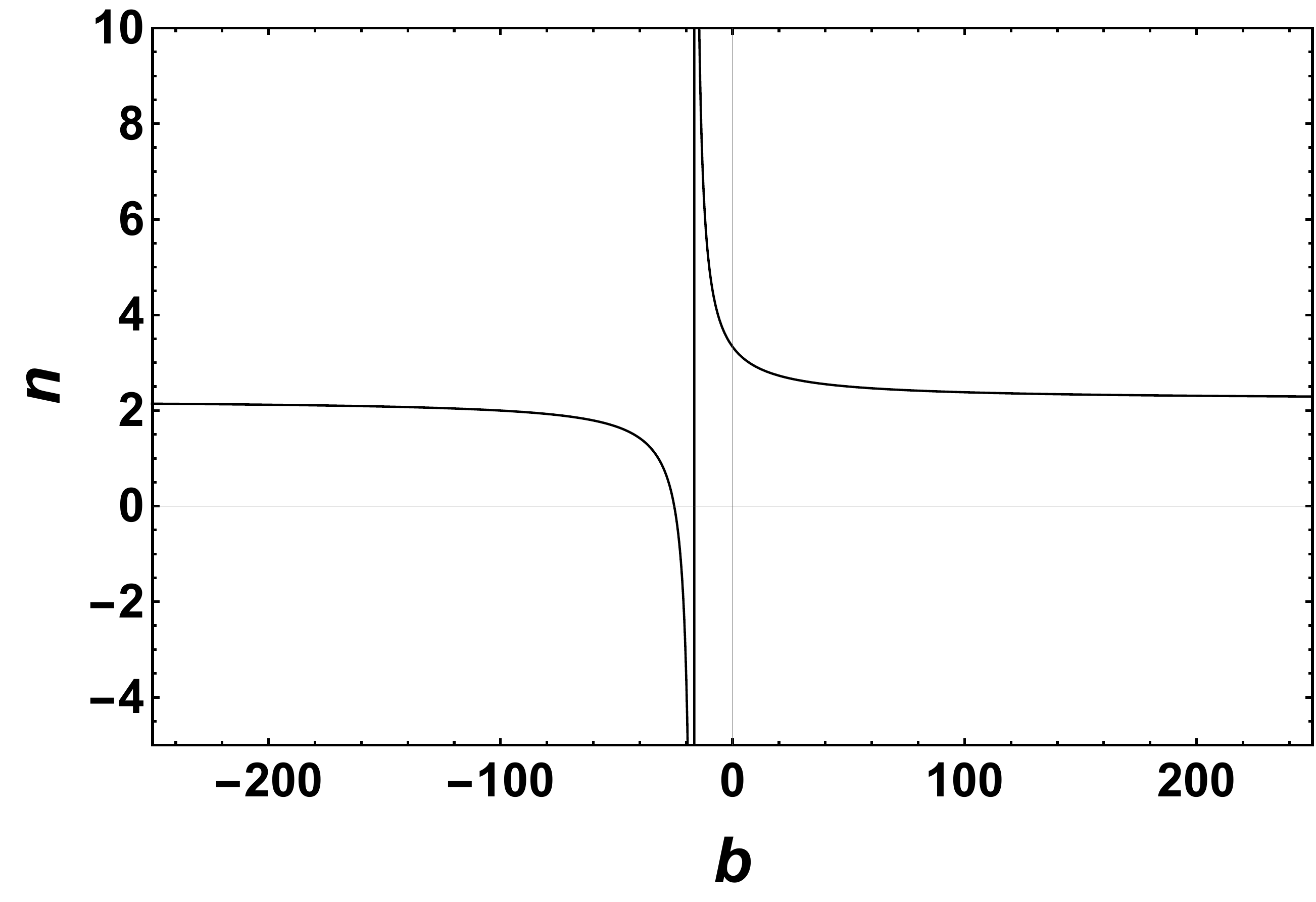} $%
\end{center}
\caption{The plot shows the variation of model parameters $n$ and $b$ in the
shown ranges to have a $q_{0}$ value consistent with observations.}
\label{n-b.pdf}
\end{figure}

In the above expressions, we can see that the Eqs. (\ref{Hz}) and (\ref{q})
have two model parameters regulating the dynamics of the model. The model
parameter $n$ is more significant than $b$ which can be analyzed in the term 
$\dfrac{24\pi +3b}{16\pi +3b}$ containing homogeneous term $3b$ in both
numerator and denominator. The choice of model parameters $n$ \& $b$ must be
in such a way that the deceleration parameter must attain a negative value
at present and consistent with the observational value of $q_{0}\simeq -0.54$
\cite{Almada/2020,Garza/2019,Akarsu/2019}. This require $n=\frac{25.142+b}{%
7.542+0.45b}$ and for a model consistent with these observations and the $%
q_{0}$ value in the neighborhood of $-0.54$, the model parameters $n$ and $b$
must satisfy the relation,

\begin{equation}  \label{nb}
n=\frac{25.142+b}{7.542+0.45b}\text{.}
\end{equation}
The following plot is an illustration for the choice of these two model
parameters shown in Fig. \ref{n-b.pdf}.

From the above Fig. \ref{n-b.pdf}, we can make a\ rough estimate for the
range of the model parameters. The best estimate of these could be in the
range $n\in (1,3)$ and with $b\in (0,2)$. In order to obtain better
constrained values of these model parameters $n$ \& $b$, we have here
considered two observational data sets with $57$ points of Hubble data sets
and $580$ points of union 2.1 compilation supernovae data sets. The
methodology to this is explained below.

\subsection{Hubble Datasets}

\qquad In a recent paper, Sharov and Vasiliev \cite{sharov} has compiled a
list of $57$ points of $H(z)$ values in the redshift range $0.07\leqslant
z\leqslant 2.42$ (see the Table-1 with $31$ points from DA method and $26$
points from BAO \& other methods with errors).

\begin{center}
\begin{tabular}{|c|c|c|c|c|c|c|c|c|c|c|c|c|c|c|c|}
\hline
\multicolumn{16}{|c|}{Table-1: $H(z)$ datasets} \\ \hline
$z$ & $H(z)$ & $\sigma _{H}$ & Ref. & $z$ & $H(z)$ & $\sigma _{H}$ & Ref. & $%
z$ & $H(z)$ & $\sigma _{H}$ & Ref. & $z$ & $H(z)$ & $\sigma _{H}$ & Ref. \\ 
\hline
$0.070$ & $69$ & $19.6$ & \cite{h14} & $0.4783$ & $80.9$ & $9$ & \cite{h18}
& $0.24$ & $79.69$ & $2.99$ & \cite{h7} & $0.52$ & $94.35$ & $2.64$ & \cite%
{h6} \\ \hline
$0.90$ & $69$ & $12$ & \cite{h13} & $0.480$ & $97$ & $62$ & \cite{h14} & $%
0.30$ & $81.7$ & $6.22$ & \cite{h10} & $0.56$ & $93.34$ & $2.3$ & \cite{h6}
\\ \hline
$0.120$ & $68.6$ & $26.2$ & \cite{h14} & $0.593$ & $104$ & $13$ & \cite{h16}
& $0.31$ & $78.18$ & $4.74$ & \cite{h6} & $0.57$ & $87.6$ & $7.8$ & \cite{h2}
\\ \hline
$0.170$ & $83$ & $8$ & \cite{h13} & $0.6797$ & $92$ & $8$ & \cite{h16} & $%
0.34$ & $83.8$ & $3.66$ & \cite{h7} & $0.57$ & $96.8$ & $3.4$ & \cite{h5} \\ 
\hline
$0.1791$ & $75$ & $4$ & \cite{h16} & $0.7812$ & $105$ & $12$ & \cite{h16} & $%
0.35$ & $82.7$ & $9.1$ & \cite{h1} & $0.59$ & $98.48$ & $3.18$ & \cite{h6}
\\ \hline
$0.1993$ & $75$ & $5$ & \cite{h16} & $0.8754$ & $125$ & $17$ & \cite{h16} & $%
0.36$ & $79.94$ & $3.38$ & \cite{h6} & $0.60$ & $87.9$ & $6.1$ & \cite{h8}
\\ \hline
$0.200$ & $72.9$ & $29.6$ & \cite{h15} & $0.880$ & $90$ & $40$ & \cite{h14}
& $0.38$ & $81.5$ & $1.9$ & \cite{h11} & $0.61$ & $97.3$ & $2.1$ & \cite{h11}
\\ \hline
$0.270$ & $77$ & $14$ & \cite{h13} & $0.900$ & $117$ & $23$ & \cite{h13} & $%
0.40$ & $82.04$ & $2.03$ & \cite{h6} & $0.64$ & $98.82$ & $2.98$ & \cite{h6}
\\ \hline
$0.280$ & $88.8$ & $36.6$ & \cite{h15} & $1.037$ & $154$ & $20$ & \cite{h16}
& $0.43$ & $86.45$ & $3.97$ & \cite{h7} & $0.73$ & $97.3$ & $7.0$ & \cite{h8}
\\ \hline
$0.3519$ & $83$ & $14$ & \cite{h16} & $1.300$ & $168$ & $17$ & \cite{h13} & $%
0.44$ & $82.6$ & $7.8$ & \cite{h8} & $2.30$ & $224$ & $8.6$ & \cite{h9} \\ 
\hline
$0.3802$ & $83$ & $13.5$ & \cite{h18} & $1.363$ & $160$ & $33.6$ & \cite{h17}
& $0.44$ & $84.81$ & $1.83$ & \cite{h6} & $2.33$ & $224$ & $8$ & \cite{h12}
\\ \hline
$0.400$ & $95$ & $17$ & \cite{h13} & $1.430$ & $177$ & $18$ & \cite{h13} & $%
0.48$ & $87.79$ & $2.03$ & \cite{h6} & $2.34$ & $222$ & $8.5$ & \cite{h4} \\ 
\hline
$0.4004$ & $77$ & $10.2$ & \cite{h18} & $1.530$ & $140$ & $14$ & \cite{h13}
& $0.51$ & $90.4$ & $1.9$ & \cite{h11} & $2.36$ & $226$ & $9.3$ & \cite{h3}
\\ \hline
$0.4247$ & $87.1$ & $11.2$ & \cite{h18} & $1.750$ & $202$ & $40$ & \cite{h13}
&  &  &  &  &  &  &  &  \\ \hline
$0.4497$ & $92.8$ & $12.9$ & \cite{h18} & $1.965$ & $186.5$ & $50.4$ & \cite%
{h17} &  &  &  &  &  &  &  &  \\ \hline
$0.470$ & $89$ & $34$ & \cite{h19} &  &  &  &  &  &  &  &  &  &  &  &  \\ 
\hline
\end{tabular}
\end{center}

The mean values of the model parameters $n$ \& $b$ are determined by
minimizing the chi-square value by maximum likelihood analysis reads as,

\begin{equation}  \label{chihz}
\chi _{OHD}^{2}(p_{s})=\sum\limits_{i=1}^{28}\frac{%
[H_{th}(p_{s},z_{i})-H_{obs}(z_{i})]^{2}}{\sigma _{H(z_{i})}^{2}},
\end{equation}
where, $H_{th}$ and $H_{obs}$ respectively refers to the theoretical and
observed value of Hubble parameter $H$, $p_{s}$ refers to the parameter
space of the model to be constrained. Also $\sigma _{H(z_{i})}$ stands for
the standard error in the observed value of $H$.

\subsection{Supernovae datasets}

\qquad The other data sets, we use for our analysis is the Union $2.1$
compilation supernovae data sets \cite{SNeIa} with $580$ points. The
chi-square formula for the supernovae data sets is given as,

\begin{equation}  \label{chisn}
\chi _{SN}^{2}(\mu _{0},p_{s})=\sum\limits_{i=1}^{580}\frac{[\mu
_{th}(\mu_{0},p_{s},z_{i})-\mu _{obs}(z_{i})]^{2}}{\sigma _{\mu (z_{i})}^{2}}%
,
\end{equation}
where, $\mu _{th}$ and $\mu _{obs}$ are respectively, the theoretical and
observed distance modulus with the standard error in the observed value
denoted by $\sigma _{\mu (z_{i})}$. The distance modulus $\mu (z)$ is
defined by $\mu (z)=m-M=5LogD_{l}(z)+\mu _{0},$where $m$ and $M$ are
respectively, the apparent and absolute magnitudes of a standard candle and
the luminosity distance $D_{l}(z)$ and the nuisance parameter $\mu _{0}$ are
defined by $D_{l}(z)=(1+z)H_{0}\int_{0}^{z}\frac{1}{H(z^{\ast })}dz^{\ast }$
and $\mu _{0}=5Log\frac{H_{0}^{-1}}{Mpc}+25$ respectively. In order to
calculate luminosity distance, we have restricted the series of $H(z)$ up to
tenth term then integrate the approximate series to obtain the luminosity
distance.

\begin{figure}[H]
\begin{center}
$%
\begin{array}{c@{\hspace{.1in}}c}
\includegraphics[width=3.0 in, height=2.5
		in]{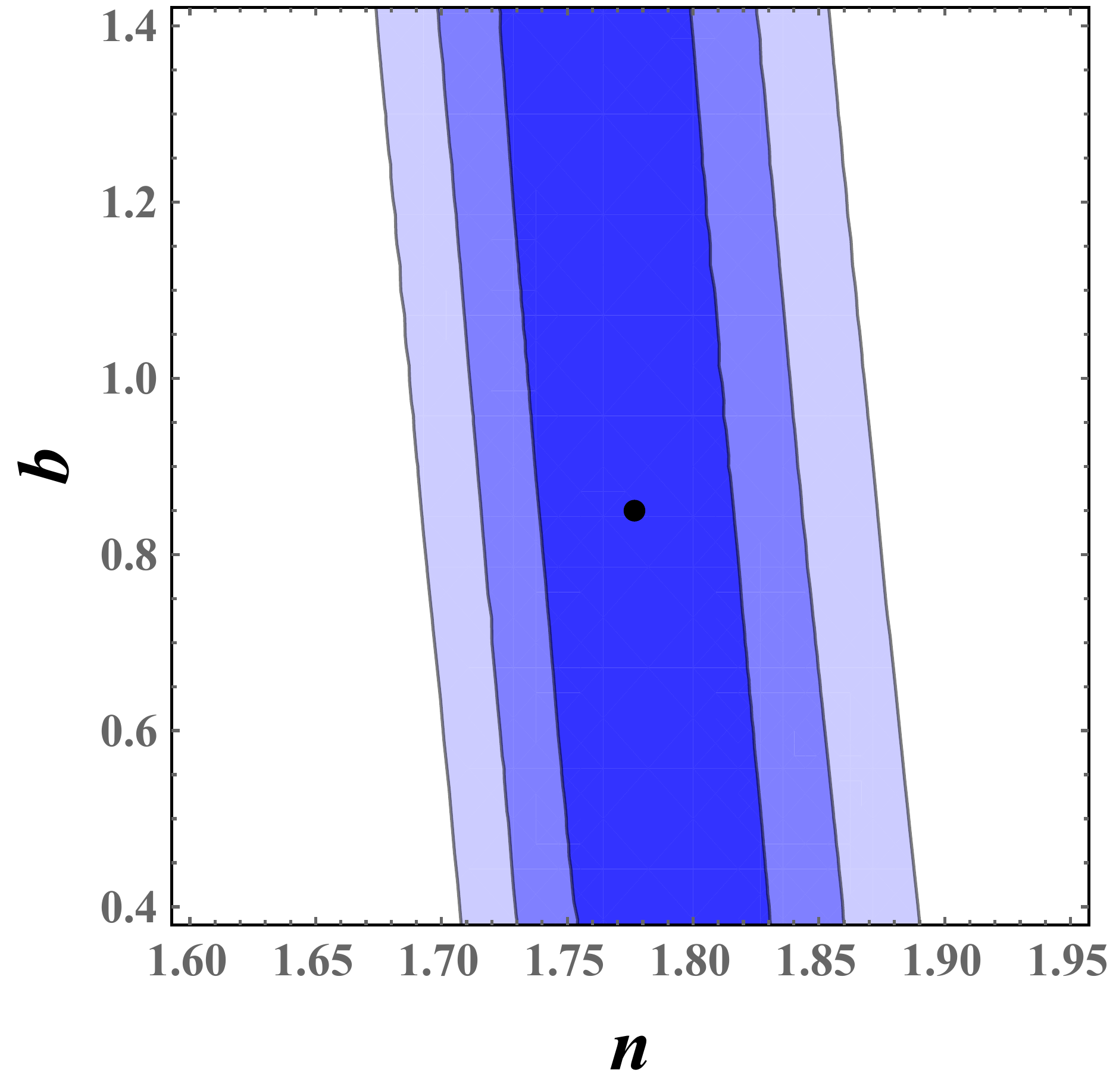} & 
\includegraphics[width=3.0 in,
		height=2.5 in]{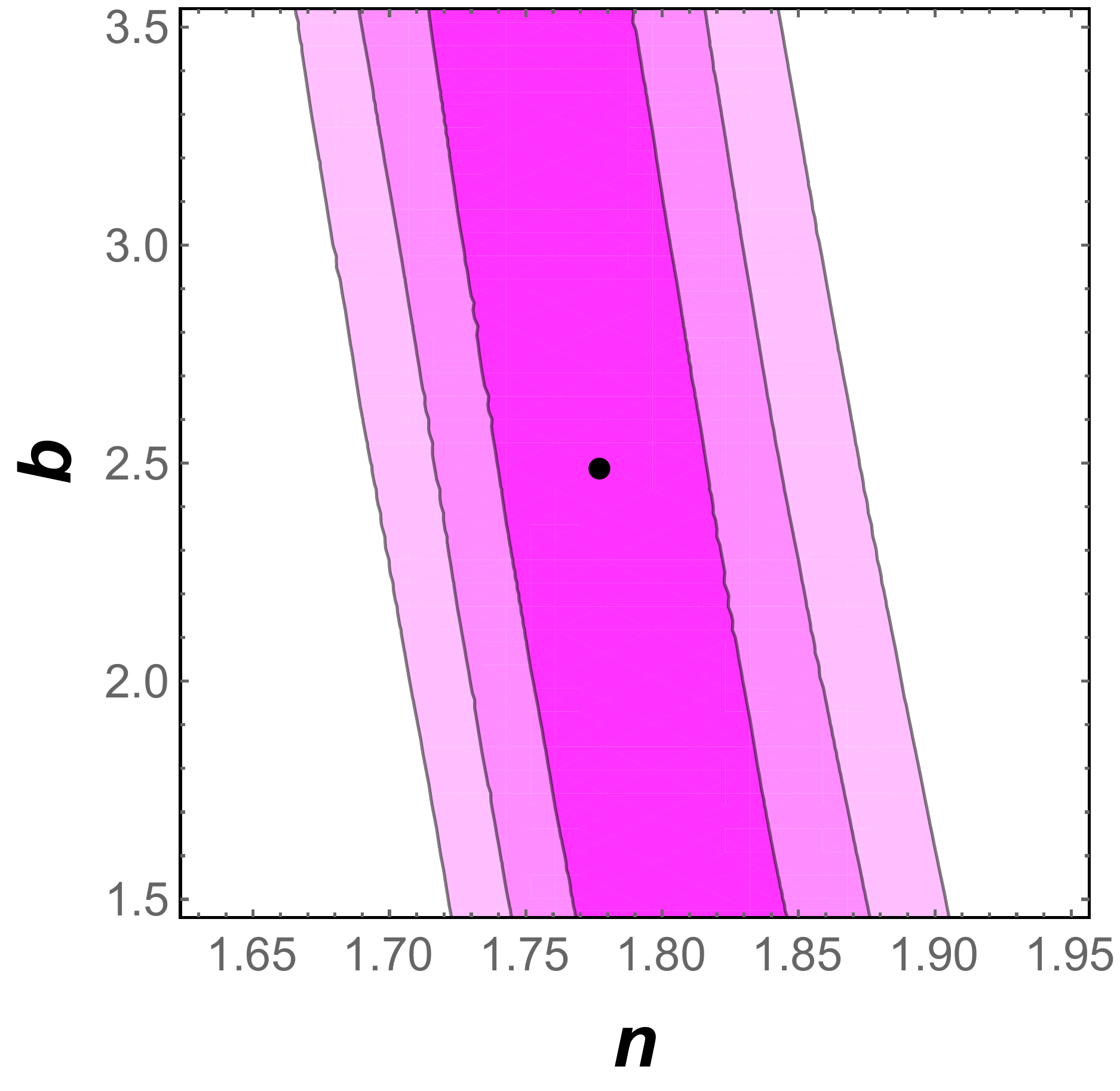} \\ 
\mbox (a) & \mbox (b)%
\end{array}
$%
\end{center}
\caption{The left panel shows the contour plot due to the $57$ points of
Hubble data sets showing likelihood values of the model parameters $n$ and $%
b $ in $n$-$b$ plane at $1\protect\sigma $, $2\protect\sigma $ \& $3\protect%
\sigma $ level and the black dot shows the constrained values of model
parameters found as $n=1.7763$ and $b=0.8491$ with $\protect\chi _{\min
}^{2}=194.3194$. The right panel shows the contour plot due to the $580$
points of union 2.1 compilation supernovae data sets showing likelihood
values of the model parameters $n$ and $b$ in $n$-$b$ plane at $1\protect%
\sigma $, $2\protect\sigma $ \& $3\protect\sigma $ level and the black dot
shows the constrained values of model parameters found as $n=1.7769$ and $%
b=2.4889$ with $\protect\chi _{\min }^{2}=790.922$.}
\label{Cont-Sn.pdf}
\end{figure}

The following plots in Fig. \ref{Cont-Sn.pdf} shows the contour plots with $%
57$ points of $H(z)$ data sets and $580$ points of $SN$ data sets showing
the likelihood values of $n$ \& $H_{0}$ at $1\sigma $, $2\sigma $ \& $%
3\sigma $ level in the $n$-$H_{0}$ plane together with the constrained
values shown in black dots.

Since, the values of the model parameter $b$ ranges from $-\infty $ to $%
+\infty $, the $b$-axis in the above contour plots are unbounded but the
model parameter $n$ is in its fixed range ($n\in (1,4)$) in both the contour
plots. We have obtained best fitting pair $(n,b)$ of model parameter values $%
(1.7763,0.8491)$ due to $H(z)$ data sets and $(1.7769,2.4889)$ due to Union
2.1 supernovae data sets. With these values, we have shown the Error bar
plots of $57$ points of $H(z)$ data sets and $580$ points of supernovae data
sets and compared our model with $\Lambda $CDM model in the following Fig. %
\ref{Error-Sn.pdf}.

\begin{figure}[H]
\begin{center}
$%
\begin{array}{c@{\hspace{.1in}}c}
\includegraphics[width=3.0 in, height=2.5
		in]{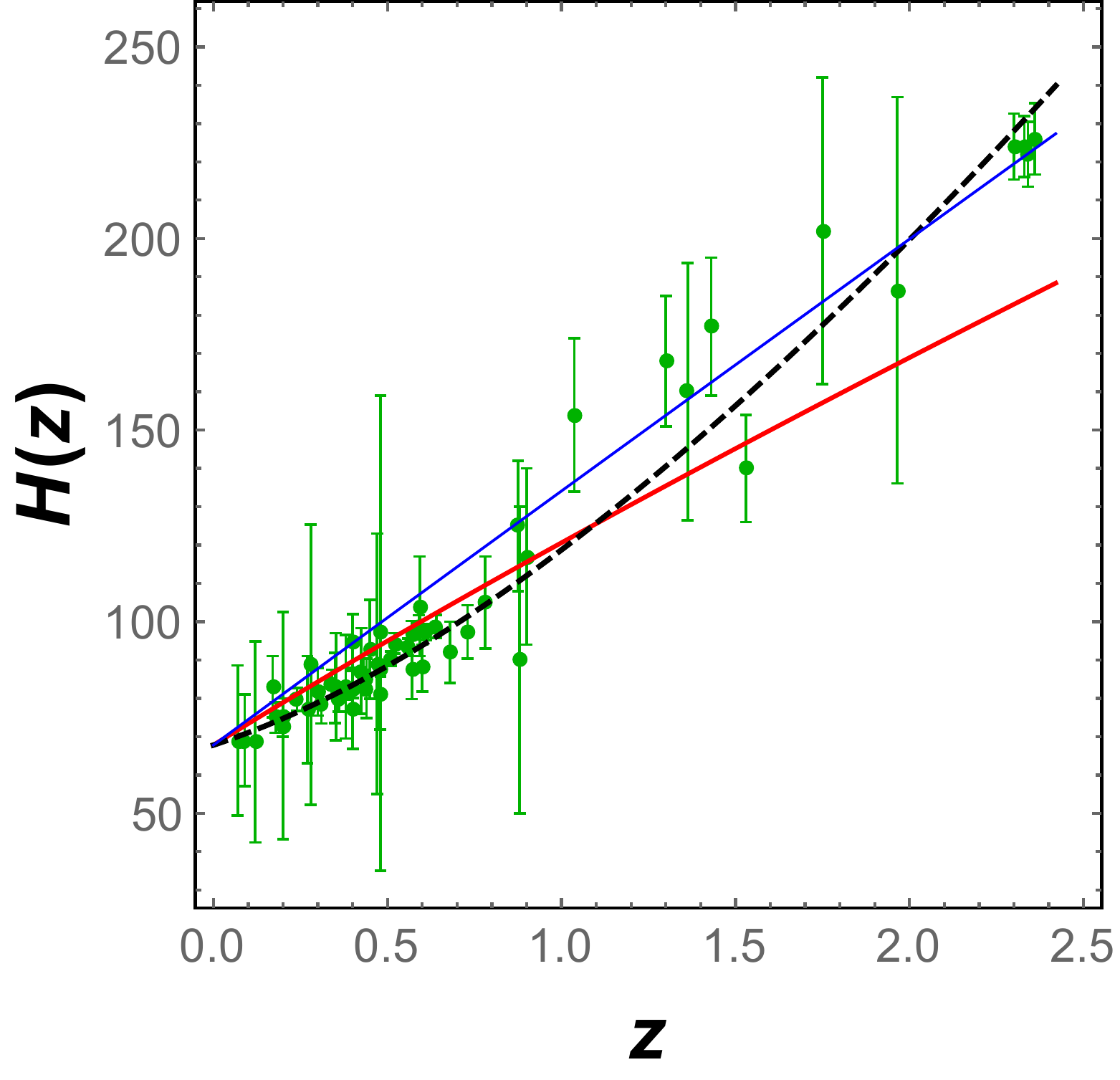} & 
\includegraphics[width=3.0 in,
		height=2.5 in]{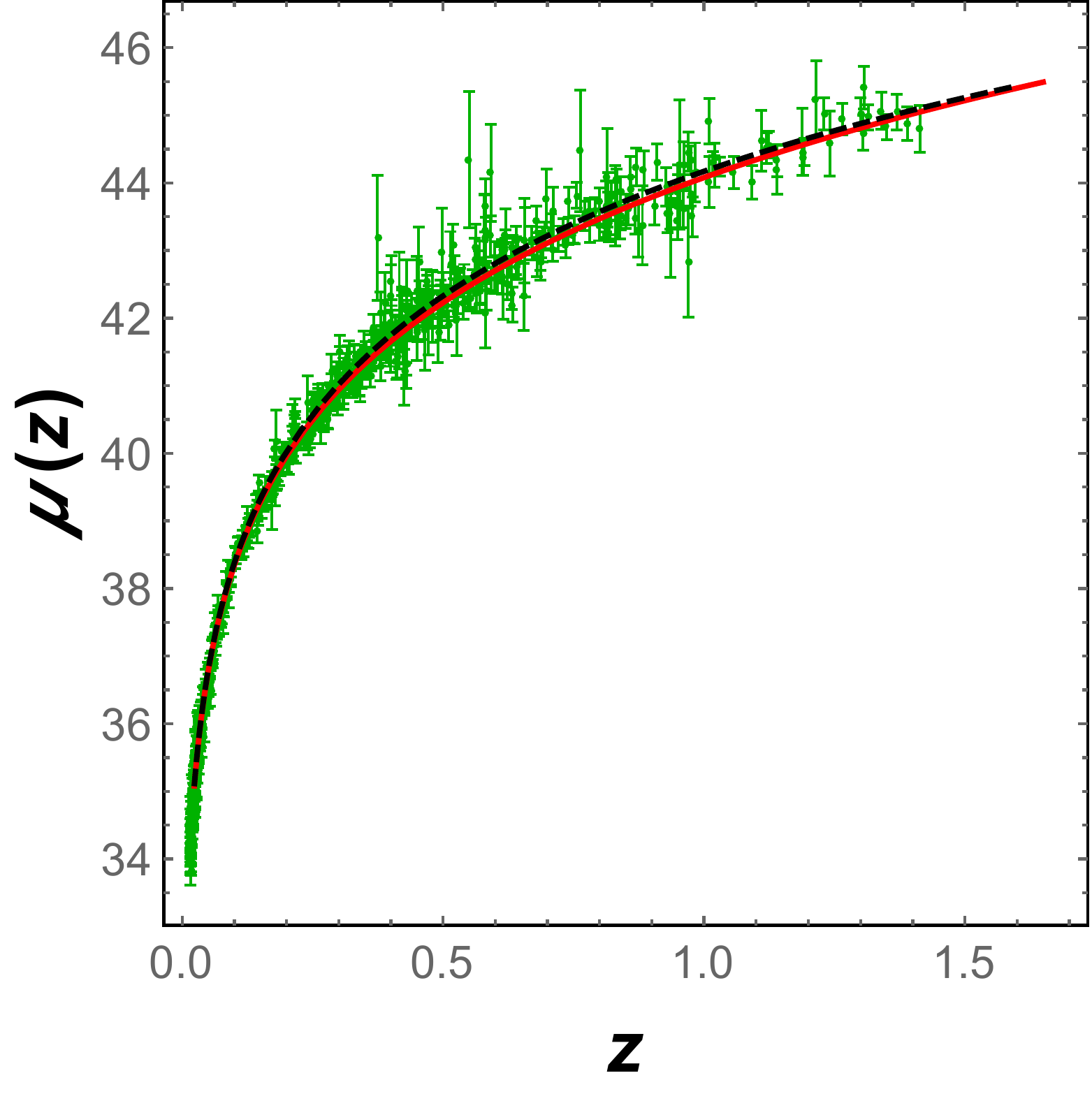} \\ 
\mbox (a) & \mbox (b)%
\end{array}
$%
\end{center}
\caption{The left panel shows the Error bar plot of $57$ points of Hubble
datasets together with the presented model shown in solid red line with $%
n=1.7767$ \& $b=0.8491$ compared with the $\Lambda $CDM model shown in black
dashed line showing a poor fit at higher redshift but better at lower
redshift. A blue line shown is a fiducial model just for comparison with the
values $n=1.5$ \& $b=0.8491$ which lies outside the contour. The right panel
shows the Error bar plot of $580$ points of union 2.1 compilation supernovae
data sets together with the presented model shown in solid red line with $%
n=1.7765$ \& $b=2.4889$ compared with the $\Lambda $CDM model shown in black
dashed line.}
\label{Error-Sn.pdf}
\end{figure}

\section{Diagnostic Analysis}\label{V}

\subsection{Statefinder diagnostic}

The well-known geometrical parameters in cosmology are Hubble parameter $H=%
\frac{\dot{a}}{a}$ and deceleration parameter $q=\frac{-\ddot{a}}{aH^{2}}$
which are useful in describing the expansion history of the Universe. Also,
various dark energy models have been proposed to explain the accelerated
expansion of the Universe. Another parameters proposed are \cite%
{Statefinder1} known as statefinder parameters written in pairs as $\{r,s\}$
and $\{r,q\}$. These are geometrical quantities engaged to identify the
various dark energy models \cite{Statefinder1, Statefinder2, Rani}. The $r$
and $s$ parameters are defined as, 
\begin{equation}
r=\frac{\dddot{a}}{aH^{3}},
\end{equation}

\begin{equation}
s=\frac{r-1}{3(q-\frac{1}{2})}.
\end{equation}

The plot of $s$-$r$ and $q$-$r$ plane is shown below in Fig. \ref%
{Statefinder.pdf}.

\begin{figure}[H]
\begin{center}
$%
\begin{array}{c@{\hspace{.1in}}c}
\includegraphics[width=3.0 in, height=2.5
		in]{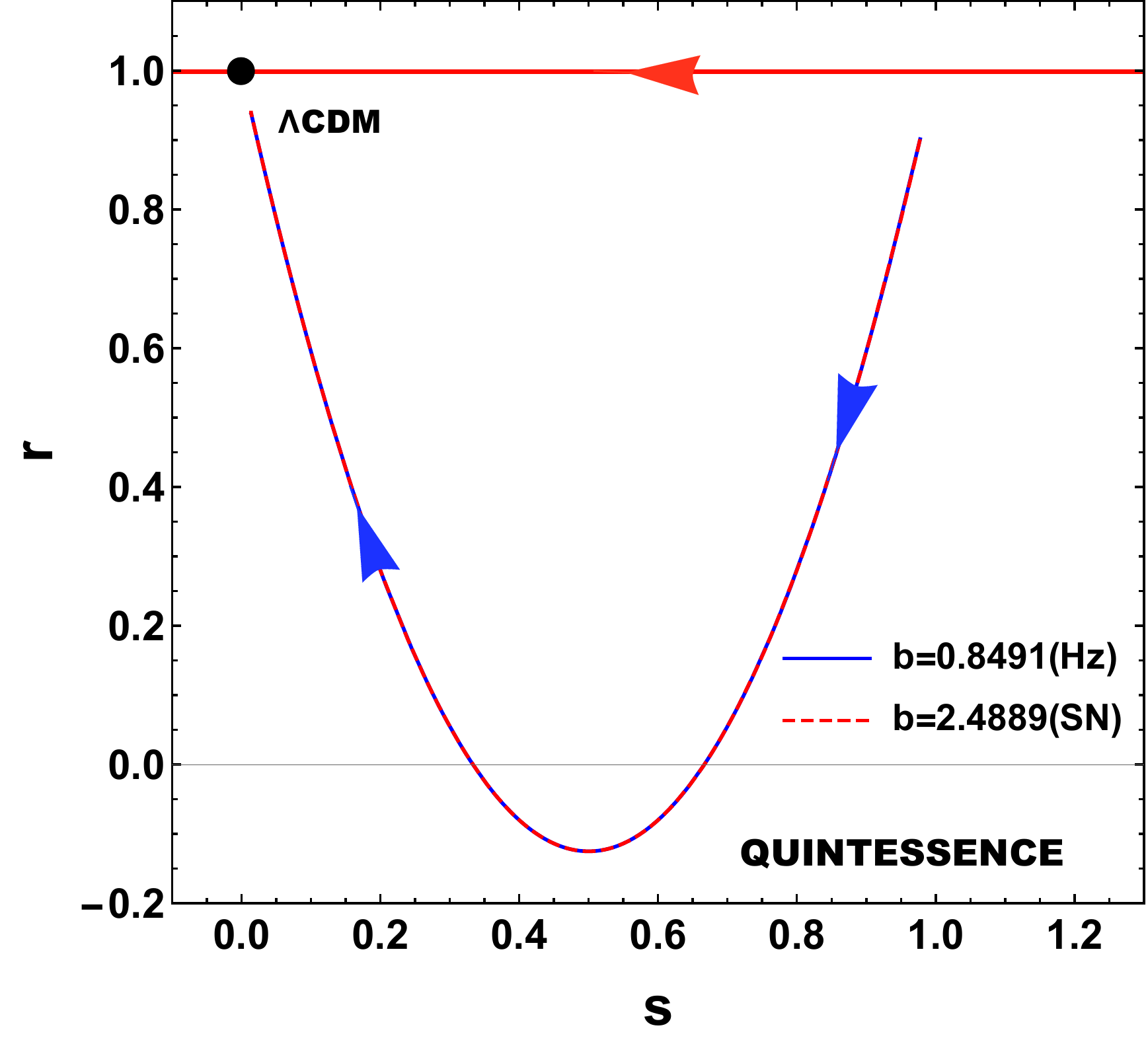} & \includegraphics[width=3.0 in,
		height=2.5 in]{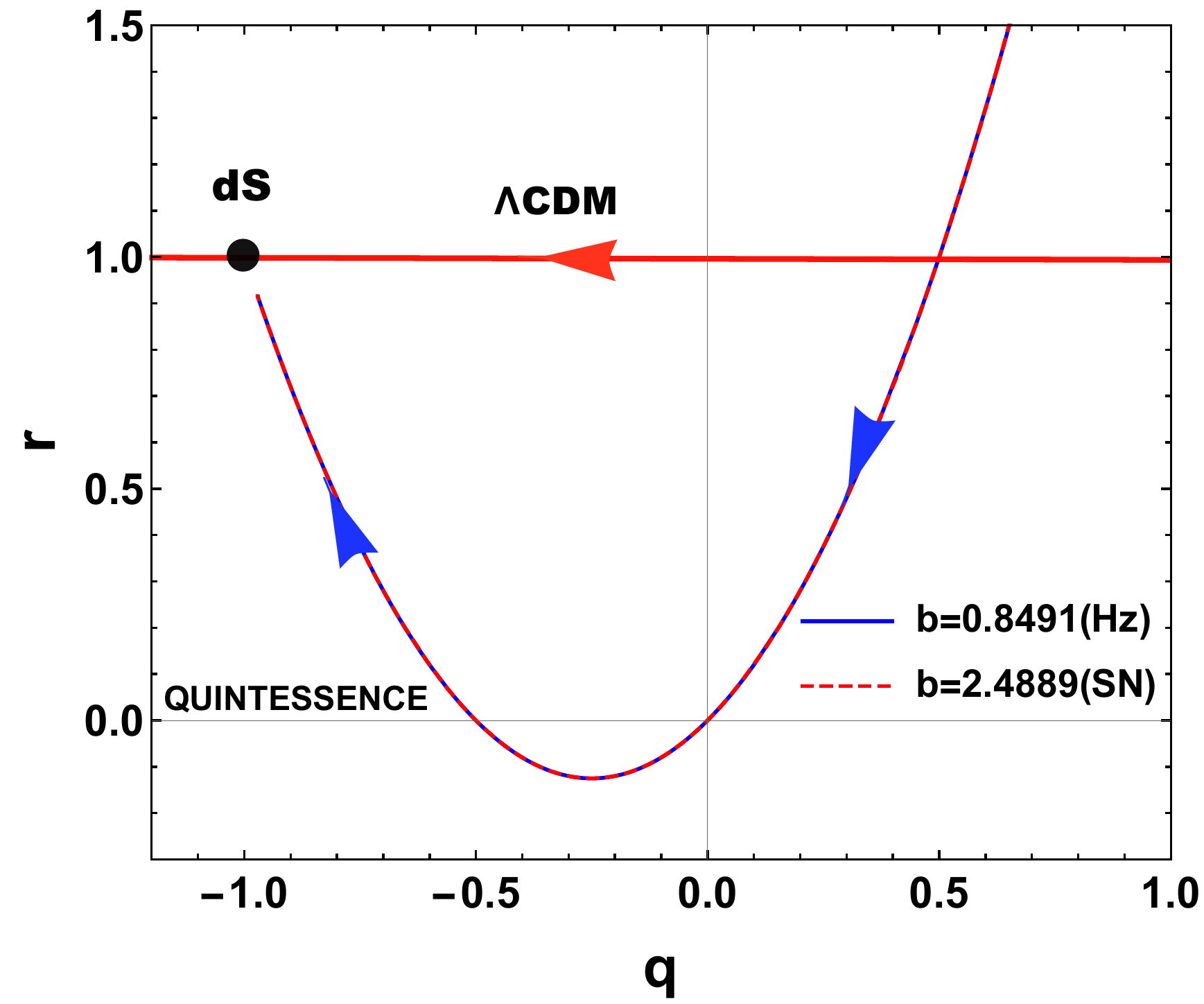} \\ 
\mbox (a) & \mbox (b)%
\end{array}
$%
\end{center}
\caption{The left panel shows the $s$-$r$ plane for our model with $b=0.8491 
$ \& $b=2.4889$ and varying $n$. The right panel shows the $q$-$r$ plane for
the model with $b=0.8491$ \& $b=2.4889$ and varying $n$.}
\label{Statefinder.pdf}
\end{figure}

Various dark energy models in $s$-$r$ plane illustrate some different
growing trajectories. The point $(s,r)=(0,1)$ correspond to the $\Lambda $%
CDM model in flat FLRW background. It is observed in Fig. \ref%
{Statefinder.pdf} that the point $(s,r)=(0,1)$ represent the $\Lambda $CDM
while $(q,r)=(-1,1)$ the de Sitter(dS) point. The red line divides the plane
into two parts denoting the Quintessence phase as a lower half. The
statefinder plots have been done for the values of n and b constrained by
the $H(z)$ and $SN$ data sets. The similar behavior can be seen in \cite%
{Rani, Suresh}. We note that for the Hubble data sets, the $r$ and $s$
parameters at the present epoch are $r_{0}=-0.111918$ and $s_{0}=0.553917$
while for the $SN$ data sets, $r_{0}=-0.118324$ and $s_{0}=0.538516$.
Currently observations are not sensitive enough to measure these parameters.
However, Ref \cite{rso} claims that these parameters can be deduced from
future observations which would greatly help to constrain the nature of dark
energy.

\subsection{$Om$ diagnostic}

$Om$ diagnostic is another efficient tool emerging from Hubble parameter
which provides the null test for the $\Lambda $CDM model. The diagnostic is
adequate in refining various dark energy models from $\Lambda $CDM due to
the variation in its slope. For flat Universe, the Om(z) is defined as \cite%
{Omdiagnostic1, Omdiagnostic2} 
\begin{equation}
Om(z)=\dfrac{\left( \frac{H(z)}{H_{0}}\right) ^{2}-1}{(1+z)^{3}-1}.
\end{equation}

\begin{figure}[H]
\centering
\includegraphics[scale =0.4]{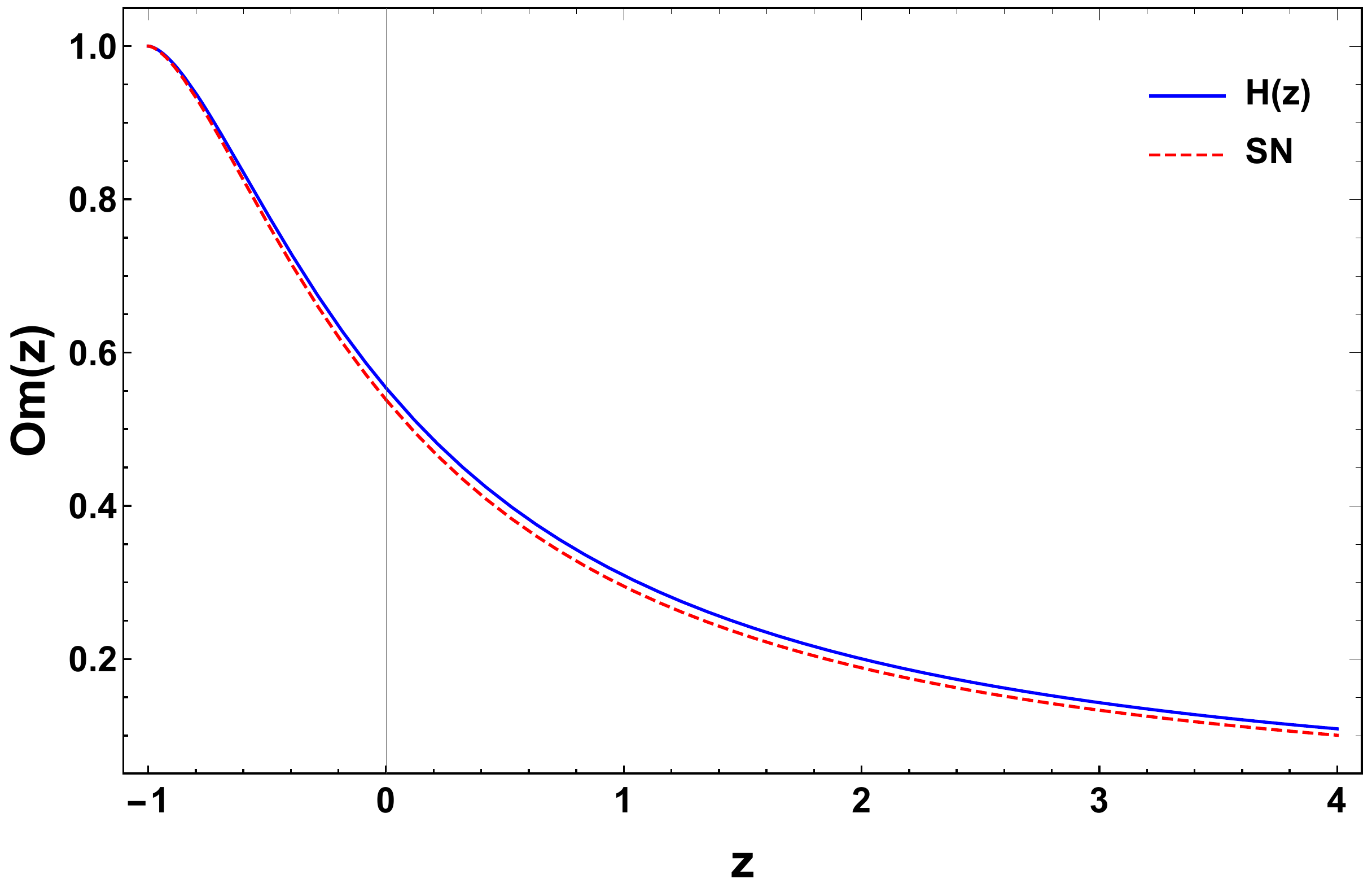}
\caption{The plot shows the behavior of $Om(z)$ with $n=1.7763$, $b=0.8491$
and $n=1.7769$, $b=2.4889$.}
\label{Om}
\end{figure}

For the $\Lambda $CDM model, phantom and quintessence cosmological models, $%
Om(z)$ have some different set of values. We can describe the behavior of
dark energy as quintessence type which shows negative curvature, phantom
type which shows positive curvature and $\Lambda $CDM the zero curvature.
Fig. \ref{Om} depicts the behavior of $Om(z)$ showing a decaying behavior
for the constrained values of the model parameters obtained from $H(z)$ and $%
SN$ data sets.

\section{Evolution of the $\protect\rho (z)$}\label{VI}

The expression for the energy density in Eq. (\ref{13}) can be written in
terms of redshift $z$ as, 
\begin{equation}
\rho (z)=\dfrac{6^{n}m(1-2n)}{16\pi +3b}H_{0}^{2n}(1+z)^{2\left( \frac{24\pi
+3b}{16\pi +3b}\right) }\text{.}  \label{d1}
\end{equation}

\begin{figure}[H]
\begin{center}
$%
\begin{array}{c@{\hspace{.1in}}c}
\includegraphics[width=3.0 in, height=2.5
		in]{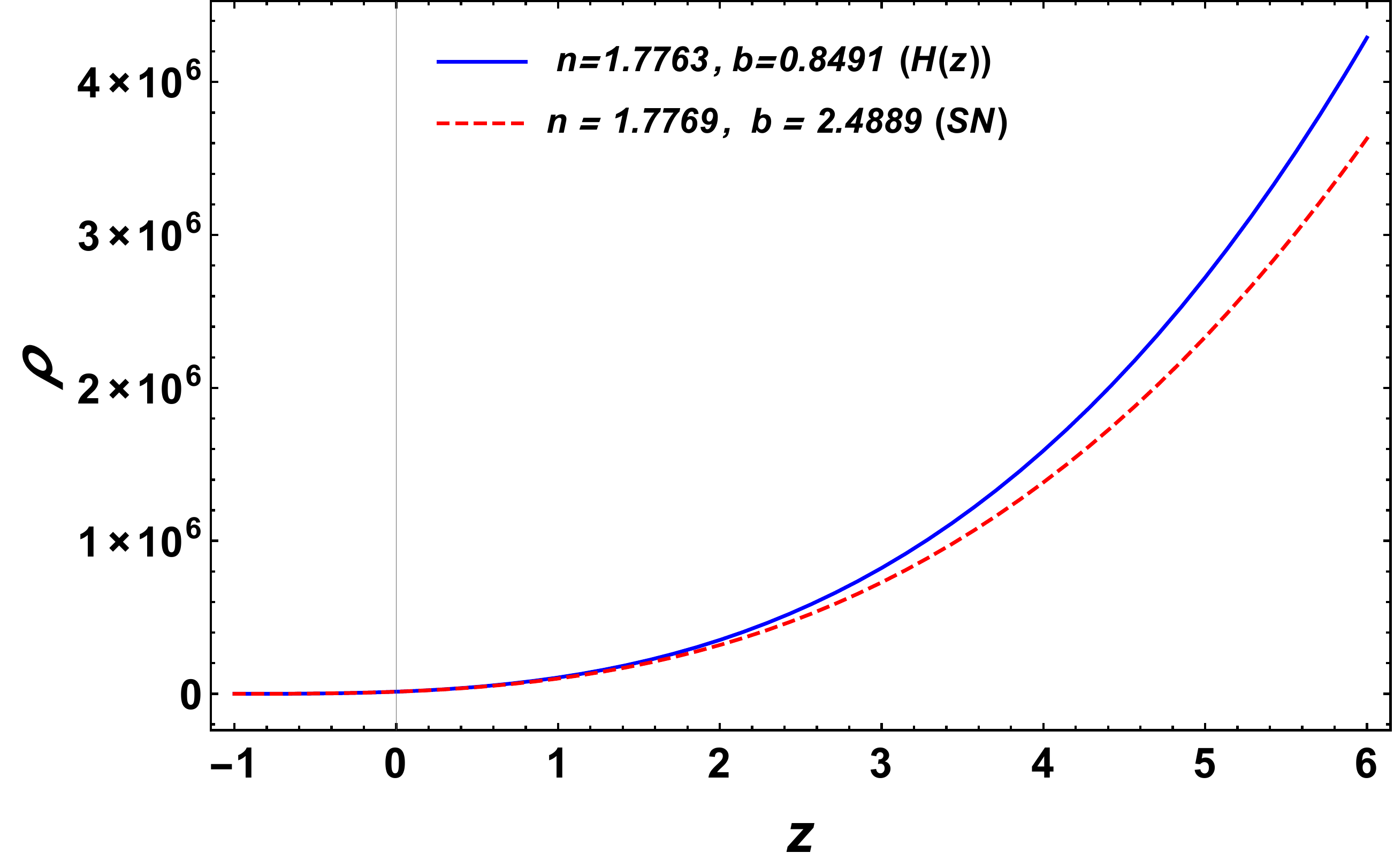} & 
\includegraphics[width=3.0 in,
		height=2.5 in]{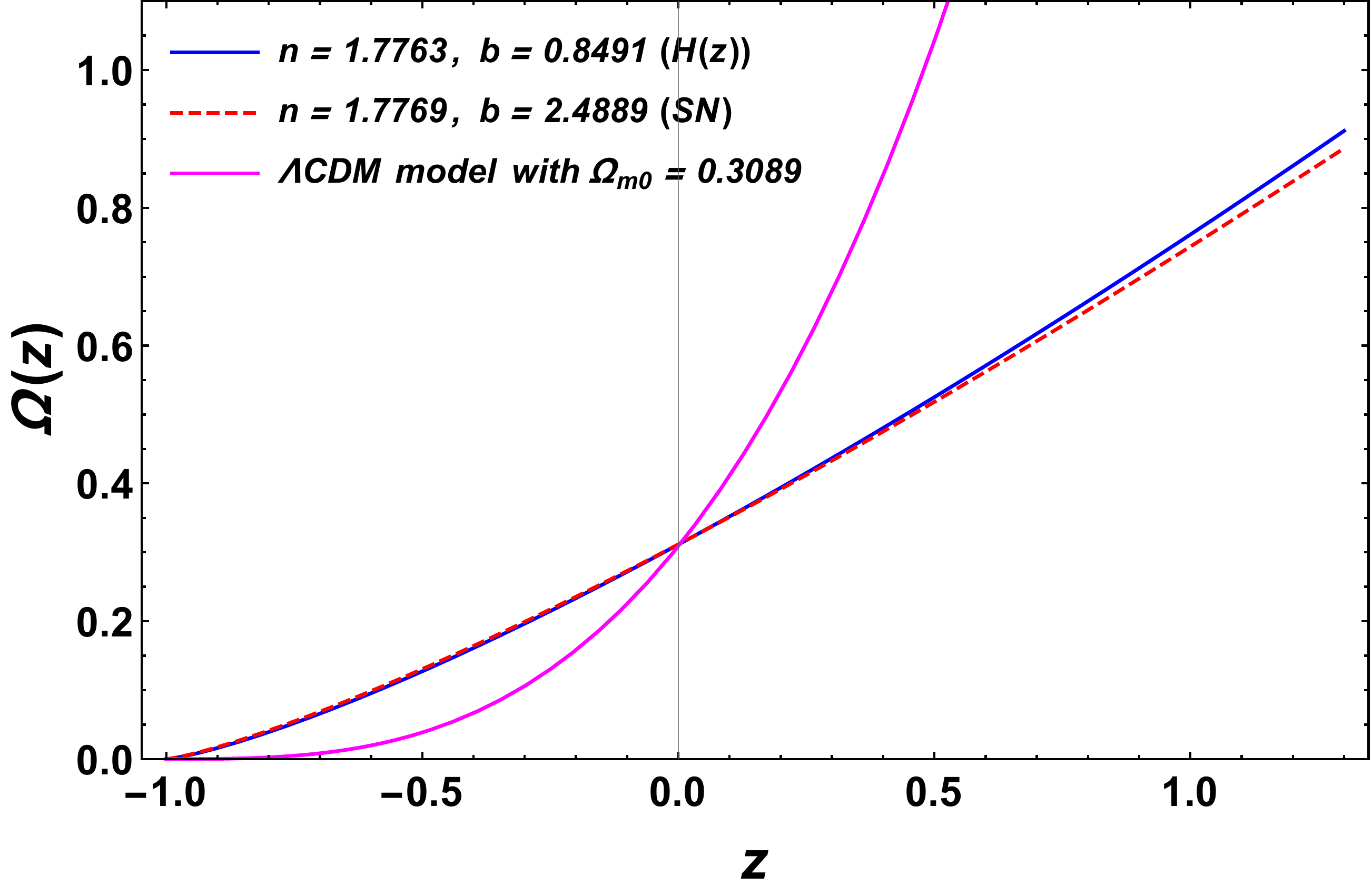} \\ 
\mbox (a) & \mbox (b)%
\end{array}
$%
\end{center}
\caption{TThe plots show the evolution of energy density (left panel) and the
density parameter (right panel) for the obtained model versus redshift $z$
with $m=-0.00115\ $ for H(z) \& $m=-0.00125\ $ for SN data constrained
values of model parameters of $n\ \&\ b$ together with $H_{0}=67.8$ km/s/Mpc. The $%
\Lambda $CDM model matter density behavior is also shown for comparision
with $\Omega _{m0}=0.3089$.}
\label{omegaR}
\end{figure}

Now, defining the density parameter, $\Omega =\frac{8\pi G\rho }{3H^{2}}$
for which we have, 
\begin{equation}
\Omega (z)=\frac{8\pi G}{3\left( 16\pi +3b\right) }\left[ 6^{n}m(1-2n)\right]
H_{0}^{2n-2}(1+z)^{2\left( \frac{n-1}{n}\right) \left( \frac{24\pi +3b}{%
16\pi +3b}\right) },  \label{d2}
\end{equation}%
with $\Omega (0)=\frac{8\pi G}{3\left( 16\pi +3b\right) }\left[ 6^{n}m(1-2n)%
\right] H_{0}^{2n-2}$. Since, the expressions in Eq. (\ref{d1}) and Eq. (\ref%
{d2}), there's a term $(1-2n)$ which will be negative for the discussed
range of values of $n$, we must take the adjustable free parameter $m$ in
these expressions so that $\rho $ assumes positive value. We, have shown the
evolution of the energy density with respect to redshift $z$ for the
constrained numerical values of the model parameters $n$ and $b$ in the Fig. %
\ref{omegaR} (a). Also, we have shown the evolution of the density
parameter for our model with the constrained numerical values of the model
parameters $n$ and $b$ in the Fig. \ref{omegaR} (b) together with
the evolution of density parameter of matter density $\Omega
_{m}=0.3089(1+z)^{3}$ as in the $\Lambda $CDM model $H(z)=\sqrt{\Omega
_{m}(1+z)^{3}+\Omega _{\Lambda }}$ for comparision (where $\Omega
_{m0}=0.3089$ and $\Omega _{\Lambda }=0.6911$ as suggested by Planck 2015
results \cite{Hz-Plank}).

\section{Linear form of $f(Q,T)=mQ+bT$}\label{VII}

We can see for $n=1$, in the considered functional form of $f(Q,T)=mQ^{n}+bT$%
, the case reduce to the linear form of $f(Q,T)$ function as $f(Q,T)=mQ+bT$
which is similar to the linear functional form considered in \cite{Yixin}
with $f(Q,T)=\alpha Q+\beta T$.

So, for $n=1$, we can solve Eq. (\ref{11}) which yield the expression of the
density $\rho $ as, 
\begin{equation}
\rho =\frac{-6mH^{2}}{16\pi +3b}\text{.}  \label{r1}
\end{equation}%
Now from Eq.(\ref{10}), we can obtain the dynamical equation in $H$ reads
as, 
\begin{equation}
\dot{H}+\frac{3(8\pi +b)}{16\pi +3b}H^{2}=0\text{,}
\end{equation}%
which yield a Hubble parameter in the form, 
\begin{equation}
H=\frac{1}{Bt+k_{1}}  \label{H1}
\end{equation}%
where $B=\frac{24\pi +3b}{16\pi +3b}$ with $k_{1}$ a constant of
integration. From Eq. (\ref{H1}), the scale factor can be obtained as 
\begin{equation}
a(t)=k_{2}(Bt+k_{1})^{\frac{1}{B}}.  \label{a1}
\end{equation}%
where $k_{2}$ is another constant of integration. As discussed above, we
must express all the above cosmological parameters in terms of redshift $z$,
for which the $t$-$z$ relationship will be established in this case as,

\begin{equation}
t(z)=-\frac{k_{1}}{B}+\frac{1}{B}\left[ k_{2}(1+z)\right] ^{-B}.
\end{equation}

So, the Hubble parameter can be written in this case as, 
\begin{equation}
H(z)=H_{0}(1+z)^{B}=H_{0}(1+z)^{\left( \frac{24\pi +3b}{16\pi +3b}\right) }%
\text{,}  \label{Hz1}
\end{equation}%
containing only one model parameter $b$ with $q(t)=-1+\frac{24\pi +3b}{16\pi
+3b}$ which is also a constant with only one model parameter $b$. Since, the
term $\frac{24\pi +3b}{16\pi +3b}$ assumes a constant value $\simeq \frac{3}{%
2}$ for any values of $b\in (-\infty ,\infty )$, we have $q(t)=0.5$ showing
a constant deceleration and the model reduces to the standard lore with $%
H(z)=H_{0}(1+z)^{\frac{3}{2}}$ and $a(t)\varpropto \left( \frac{3}{2}%
t+k_{1}\right) ^{\frac{2}{3}}$. We note that, in this case the energy
density $\rho (z)=\frac{-6m}{16\pi +3b}H_{0}^{2}(1+z)^{3}$ where the
adjustable parameter $m$ is considered to be negative.

\section{Conclusion}\label{VIII}

In this work, we have discussed late time cosmology employing a well
motivated $f(Q,T)$ gravity model with the functional form $f(Q,T)=mQ^{n}+bT$%
, where $m,n$ and $b$ are model parameters proposed in \cite{Yixin}. By
constraining the free parameters using observational data sets of the
updated 57 points of Hubble data sets and 580 points of union 2.1
compilation supernovae data sets, we find the deceleration parameter to be
negative and reads respectively $q_{0}=-0.169125$ and $q_{0}=-0.192226$ and
therefore consistent with the present scenario of an accelerating universe.
Previous works in power law cosmology also reported similar constraints (see
for example \cite{Rani,Suresh}). For the model considered in \cite{Yixin}
with the $f(Q,T)$ function $f(Q,T)=-\gamma Q-\delta T^{2}$ ($\gamma $ and $%
\delta $ are model parameters), the solution and data analysis have already
been discussed. For another model considered in the same paper \cite{Yixin}
with the $f(Q,T)$ function $f(Q,T)=\alpha Q+\beta T$ ($\alpha $ and $\beta $
are model parameters) which is similar to the linear case with $n=1$ in our
considered $f(Q,T)=mQ^{n}+bT$ form i.e. $mQ+bT$. In this linear case, the
solution mimic the power law expansion model with $a(t)\varpropto
(Bt+c_{1})^{\frac{1}{B}}$ where $B=\frac{24\pi +3b}{16\pi +3b}$ containing
only one model parameter $b$ wherein we can see the parameter $b$ contribute
very less in the evolution because the term $B=\frac{24\pi +3b}{16\pi +3b}%
\approx \frac{3}{2}$ for $b\in (-\infty ,\infty )$ implying the model
behaves similar to the standard lore of $a(t)\sim t^{\frac{2}{3}}$ with a
constant deceleration $q=0.5$.

We have thoroughly investigated the nature of dark energy mimicked by the
parametrization $f(Q,T)=mQ^{n}+bT$ with the assistance of statefinder
diagnostic in $\{s,r\}$ and $\{q,r\}$ planes and also performed the $Om$
-diagnostic analysis for the model. For the numerical values of the model
parameters $n$ and $b$ obtained from constraining our model through $57$
points of $H(z)$ data sets gives the statefinder parameters values as $%
r_{0}=-0.111918$ and $s_{0}=0.553917$ while for the numerical values of the
model parameters $n$ and $b$ obtained from constraining our model through $%
580$ points for the Union 2.1 compilation data sets gives $r_{0}=-0.118324$
and $s_{0}=0.538516$ as obtained earlier in \cite{Rani,Suresh}. We can
conclude that, the model considered here is good in explaining at present
observations but may not explain the early evolution well (as the model is
not consistent with the constraints coming from BAO data sets (not done
here)). One of the significant observation is that the model parameter $b$
which is the coefficient of the trace $T$ in the $f(Q,T)=mQ^{n}+bT$ form
considered contributes very very less in the evolution as it can be seen
from the figures Fig. \ref{Statefinder.pdf} which means linear trace $T$ do
not affect the evolution. The behavior of energy density and the density parameter with respect to redshift $z$ for the constrained values of the model parameters $n$ and $b$ are depicted in Fig. \ref{omegaR} with $\Omega _{m0}=0.3089$. Some more functional form of $f(Q,T)$ could be
explored in the same way and is deferred to our future works.

\section*{Acknowledgments}

S. A. acknowledges CSIR, Govt. of India, New Delhi, for awarding Junior
Research Fellowship. PKS acknowledges CSIR, New Delhi, India for financial
support to carry out the Research project [No.03(1454)/19/EMR-II
Dt.02/08/2019]. We are very much grateful to the honorable referee and the
editor for the illuminating suggestions that have significantly improved our
work in terms of research quality and presentation.

\end{document}